\documentstyle[epsfig]{article}
\begin{document}
\voffset .3in

\def\ie{{\it i.e.}}
\def\etal{{\it et al.}}
\def\bold#1{\setbox0=\hbox{$#1$}%
     \kern-.025em\copy0\kern-\wd0
     \kern.05em\copy0\kern-\wd0
     \kern-.025em\raise.0433em\box0 }
\makeatletter
\def\lsim{\mathrel{\mathpalette\@versim<}}
\def\gsim{\mathrel{\mathpalette\@versim>}}
\def\@versim#1#2{\vcenter{\offinterlineskip
    \ialign{$\m@th#1\hfil##\hfil$\crcr#2\crcr\sim\crcr } }}
\makeatother

\def\NPB#1#2#3{{\sl Nucl.~Phys.} {\bf{B#1}} (19#2) #3}
\def\PLB#1#2#3{{\sl Phys.~Lett.} {\bf{B#1}} (19#2) #3}
\def\PRD#1#2#3{{\sl Phys.~Rev.} {\bf{D#1}} (19#2) #3}
\def\PRL#1#2#3{{\sl Phys.~Rev.~Lett.} {\bf{#1}} (19#2) #3}
\def\ZPC#1#2#3{{\sl Z.~Phys.} {\bf C#1} (19#2) #3}
\def\PTP#1#2#3{{\sl Prog.~Theor.~Phys.} {\bf#1}  (19#2) #3}
\def\MPL#1#2#3{{\sl Mod.~Phys.~Lett.} {\bf#1} (19#2) #3}
\def\PR#1#2#3{{\sl Phys.~Rep.} {\bf#1} (19#2) #3}
\def\RMP#1#2#3{{\sl Rev.~Mod.~Phys.} {\bf#1} (19#2) #3}
\def\HPA#1#2#3{{\sl Helv.~Phys.~Acta} {\bf#1} (19#2) #3}
\relax
\topmargin-2.5cm
\begin{flushright}
\large {CERN-TH/96-08\\
SCIPP 96/06 \\
hep--ph/9601331} \\
\end{flushright}
\vskip 1.0in
\begin{center}
{\LARGE\bf Supersymmetric Hints from Precision Electroweak \\[6pt]
Data?}    \\
\vskip .5in
{\large 
{\bf Howard E. Haber}  \\
\vskip .35in
CERN, TH-Division  \\
CH--1211 Geneva 23, Switzerland \\[2pt]
 and \\[2pt]
Santa Cruz Institute for Particle Physics  \\
University of California,
Santa Cruz, CA 94064  USA}\\
\end{center}
\vskip1.75cm
\begin{center}
{\large \bf Abstract}
\end{center}
\begin{quote}
{\large
The Standard Model does not provide a very good fit to
the most recent precision electroweak data from LEP, due 
primarily to the observed branching ratios
for $Z$ decay to $b\bar b$ and $c\bar c$.  The possibility that an
extension of the Standard Model with low-energy supersymmetry can
improve the agreement between data and theory is considered.}
\end{quote}
\vfill
\begin{center}
{\large Invited Talk at the  \\   1995 International
Europhysics Conference on High Energy Physics \\
Brussels, Belgium, 27 July--2 August, 1995}\\
\end{center}
\vskip1.5cm
\begin{flushleft}
\large{CERN-TH-96-08\\
January, 1996}\\
\end{flushleft}
\setcounter{page}{0}
\newpage
%
\makeatletter
\input{wstwocl.sty}
\newcount\@tempcntc
\def\@citex[#1]#2{\if@filesw\immediate\write\@auxout{\string\citation{#2}}\fi
  \@tempcnta\z@\@tempcntb\m@ne\def\@citea{}\@cite{\@for\@citeb:=#2\do
    {\@ifundefined
       {b@\@citeb}{\@citeo\@tempcntb\m@ne\@citea\def\@citea{,}{\bf ?}\@warning
       {Citation `\@citeb' on page \thepage \space undefined}}%
    {\setbox\z@\hbox{\global\@tempcntc0\csname b@\@citeb\endcsname\relax}%
     \ifnum\@tempcntc=\z@ \@citeo\@tempcntb\m@ne
       \@citea\def\@citea{,}\hbox{\csname b@\@citeb\endcsname}%
     \else
      \advance\@tempcntb\@ne
      \ifnum\@tempcntb=\@tempcntc
      \else\advance\@tempcntb\m@ne\@citeo
      \@tempcnta\@tempcntc\@tempcntb\@tempcntc\fi\fi}}\@citeo}{#1}}
\def\@citeo{\ifnum\@tempcnta>\@tempcntb\else\@citea\def\@citea{,}%
  \ifnum\@tempcnta=\@tempcntb\the\@tempcnta\else
   {\advance\@tempcnta\@ne\ifnum\@tempcnta=\@tempcntb \else \def\@citea{-}\fi
    \advance\@tempcnta\m@ne\the\@tempcnta\@citea\the\@tempcntb}\fi\fi}
\makeatother
\setcounter{secnumdepth}{2} 

   
\title{SUPERSYMMETRIC HINTS FROM PRECISION ELECTROWEAK DATA?}

\firstauthors{Howard E. Haber}

\firstaddress{CERN, TH-Division, CH-1211 Geneva 23, Switzerland}

\secondaddress{Santa Cruz Institute for Particle Physics,
University of California, Santa Cruz, CA, 95064, USA}

\twocolumn[\maketitle
\abstracts{ 
The Standard Model does not provide a very good fit to
the most recent precision electroweak data from LEP, due 
primarily to the observed branching ratios
for $Z$ decay to $b\bar b$ and $c\bar c$.  The possibility that an
extension of the Standard Model with low-energy supersymmetry can
improve the agreement between data and theory is considered.}
]

 \pagestyle{plain}
\section{The \protect{$\bold R_b$}--%
\protect{$\bold R_c$}--\protect{$\bold\alpha_s$} crisis}

Experiments at LEP and SLC measure more than fifteen separate
electroweak observables in $Z$ decay events.  A global fit to
these observables exhibits a remarkable consistency with Standard
Model (SM) expectations, with two notable exceptions.
Defining $R_Q\equiv\Gamma(Z\to
Q\overline Q)/\Gamma(Z\to{\rm hadrons})$, with
$Q = b,c$, the LEP Electroweak Working Group global fit 
yields\cite{lepglobal}
\begin{eqnarray}
R_b=\cases{0.2219\pm 0.0017,&LEP/SLC global fit;\cr
0.2156,&SM prediction,\cr}
\label{zbbnumber1}
\end{eqnarray}
which is a $3.7\sigma$ discrepancy, and
\begin{eqnarray}
R_c=\cases{0.1543\pm 0.0074,&LEP/SLC global fit;\cr
0.1724,&SM prediction,\cr}
\label{zccnumber1}
\end{eqnarray}
which is a $2.5\sigma$ discrepancy.  Because the measurements of $R_b$ and
$R_c$ are highly correlated, it is useful to examine
the contours of $\Delta\chi^2$ in the
$R_b$--$R_c$ plane with respect to the best fit to the observed 
data.\cite{charlton}
When this is done, one finds that the Standard Model
prediction lies just outside the $99.9\%$ contour.  Taken at
face value, this would suggest that the probability that the Standard
Model describes the data is less than one in a thousand!

One other LEP measurement relevant to this discussion is the
$\alpha_s(m_Z)$ determination from the total hadronic width of the 
$Z$.  Based on the measurement of 
$R_\ell\equiv\Gamma_{\rm had}/\Gamma_{\ell\ell}$, Ref.~1 finds
$\alpha_s(m_Z)=0.126\pm 0.005\pm 0.002$ (where the last error
quoted corresponds to varying the Higgs mass from 60~GeV to 1~TeV).
The LEP determination of $\alpha_s(m_Z)$ tends to be 
somewhat higher than the extrapolated value of $\alpha_s(m_Z)$
obtained from lower energy measurements.  In a recent review for the
Particle Data group,  Hinchliffe quotes\cite{hinchliffe} 
extrapolated values of
$\alpha_s(m_Z)= 0.112\pm 0.005$ from low-energy deep
inelastic scattering data and $\alpha_s(m_Z)= 0.115\pm 0.003$ from
a lattice QCD determination based on bottomonium spectroscopy.
Shifman has argued eloquently \cite{shifman} that the tendency
of lower values of
$\alpha_s(m_Z)$ determined from low-energy observables as compared to
the higher values of $\alpha_s(m_Z)$ measured at LEP presents a
serious discrepancy that could be a signal of new physics beyond the
Standard Model.

There may be a 
connection between the $\alpha_s(m_Z)$ ``discrepancy'' and the
$R_b$ and $R_c$ measurements.\cite{blondel}  If new electroweak
physics contributes positively [negatively] to
$\Gamma_{\rm had}$, then the QCD contribution to $\Gamma_{\rm had}$
determined from LEP data must be reduced [increased], 
since the sum is fixed by the observed data.  Consequently, the
value of $\alpha_s(m_Z)$ determined at LEP from $\Gamma_{\rm had}$
would have to be reduced [increased].  Thus, 
better agreement between the value of $\alpha_s(m_Z)$ as determined
from $\Gamma_{\rm had}$ and lower energy data could be achieved
if there exists a positive contribution of new physics to $\Gamma_{\rm
had}$.  

The required magnitude of the new contribution can be
determined as follows.  
Let $\Gamma_{\rm had}^{(0)}$ be the tree-level decay rate for
$Z\to{\rm hadrons}$ in the Standard Model, and let $\alpha_s^{(0)}$ be
the value of $\alpha_s(m_Z)$ extracted from LEP data based on the
measured value of $Z\to{\rm hadrons}$ under the SM
hypothesis.  If there is a non-SM electroweak component to
$\Gamma_{\rm had}$, denoted below by $\delta\Gamma_{\rm new}$,
then the true value of $\alpha_s$ should be determined (in the
approximation where QCD effects are treated at one-loop) by
\begin{equation} \label{gammanew}
\Gamma_{\rm had}=\Gamma_{\rm had}^{(0)}\left(1+{\alpha_s^{(0)}\over\pi}
\right)=\Gamma_{\rm had}^{(0)}\left(1+{\alpha_s\over\pi}\right)+
\delta\Gamma_{\rm new}\,.
\end{equation}

As an example, suppose that new electroweak physics contributes only to
$R_b$, and not to $R_c$ or $R_q$ (where $q$ is a light quark flavor).
Then,
\begin{equation} \label{gammabbnew} 
\Gamma_{b\bar b}=\Gamma_{b\bar b}^{(0)}\left(1+{\alpha_s^{(0)}\over\pi}
\right)=\Gamma_{b\bar b}^{(0)}\left(1+{\alpha_s\over \pi}\right)+
\delta\Gamma_{\rm new}\,,
\end{equation}
where $\Gamma_{b\bar b}^{(0)}$ is the SM tree-level decay rate for
$Z\to b\bar b$.  Note that by assumption,
$\delta\Gamma_{\rm new}$ is the same quantity in eqs.~(\ref{gammanew})
and (\ref{gammabbnew}).   Let $R_b^{\rm SM}=\Gamma_{b\bar b}^{(0)}/
\Gamma_{\rm had}^{(0)}$ be the predicted value of $R_b$ in the
Standard Model (note that the dependence on $\alpha_s$ drops out
in the ratio at one-loop).  Then, 
\begin{equation}\label{arebee}
R_b={\Gamma_{b\bar b}\over \Gamma_{\rm had}}=
{\Gamma_{b\bar b}^{(0)}(1+\alpha_s/\pi)+\delta\Gamma_{\rm new}\over 
\Gamma_{\rm had}^{(0)}(1+\alpha_s/\pi)+\delta\Gamma_{\rm new}}\,.
\end{equation}
Inserting $\Gamma_{b\bar b}^{(0)}=R_b^{\rm SM}\Gamma_{\rm
had}^{(0)}$ in eq.~(\ref{arebee}), and eliminating $\delta\Gamma_{\rm new}$
using eq.~(\ref{gammanew}), all factors of $\Gamma_{\rm had}^{(0)}$ 
drop out and one can solve for $\alpha_s$.  The result is:
\begin{equation} \label{newalpha}
{\alpha_s(m_Z)\over\pi} = \left({1-R_b\over 1-R_b^{\rm SM}}\right)
\left({\alpha_s^{(0)}(m_Z)\over\pi}\right)-
 \frac{R_b - R_b^{\rm SM}}{1-R_b^{\rm SM}},
\end{equation}
As an exercise, let us insert $R_b= 0.2219$, $R_b^{\rm SM}= 0.2156$,
and $\alpha_s^{(0)}=0.126$.  Using eq.~(\ref{newalpha}), we would then find 
$\alpha_s(m_Z)= 0.100$, which is somewhat lower than any of the 
values of $\alpha_s(m_Z)$ quoted above.

In the above example, I assumed that there was no new physics
contribution to $R_c$.  Nevertheless, one should still expect a slight shift
from the SM prediction,
$R_c^{\rm SM}=\Gamma_{c\bar c}^{(0)}/\Gamma_{\rm had}^{(0)}$.
Following similar steps as above, 
\begin{equation}\label{arecee}
R_c={\Gamma_{c\bar c}\over \Gamma_{\rm had}}=
R_c^{\rm SM}\left({1+\alpha_s/\pi\over 1+\alpha_s^{(0)}/\pi}\right)\,, 
\end{equation}
from which it follows that:
\begin{equation} \label{areceenew}
R_c=R_c^{\rm SM}\left({1-R_b\over 1-R_b^{\rm SM}}\right)\,.
\end{equation}
Using the same numbers as before with $R_c^{\rm
SM}=0.172$, one would predict $R_c= 0.171$.

One can consider other scenarios.  For example, if new physics
contributes only to $R_c$, then the above formulae can be used by
interchanging $b$ and $c$ everywhere.  For $R_c= 0.1543$, one
would find 
$R_b=0.2202$.  Unfortunately, the value of $\alpha_s$
obtained is $\alpha_s(m_Z)=0.196$, which is completely inconsistent
with other measurements.  

One must be very careful in interpreting the observed $R_b$ and $R_c$
discrepancies from Standard Model expectations.
The experimental procedures that identify $b$ and $c$ quarks in
$Z$ decays are difficult and prone to large
systematic errors.  Regarding the $R_c$ measurement, note
that the quoted error is larger, and the
statistical significance of the deviation from the Standard Model
prediction is smaller than those of $R_b$.  Moreover, 
the experimentally observed
value for $R_b+R_c$ is {\it lower} than the corresponding SM
prediction.
Hence, if new physics contributes only to $R_b$
and $R_c$, then the QCD contribution
to $\Gamma_{\rm had}$ must be {\it larger} than its value in the
Standard Model, implying a value of $\alpha_s(m_Z)$ that is too large. 
Of course, this statement implicitly assumes that there are
no new physics contributions to $R_q$ where $q$
is a light quark.   However, there is no known source
of new physics that can modify $R_q$ sufficiently to compensate
the deficit in $R_b+R_c$ to avoid the above conclusion.
Thus, I am inclined to discount the measured value of $R_c$ above,
and assume that its true value is close to the Standard Model
expectation.  

Should one discount the measured value of $R_b$ as
well?  Further experimental analysis is required to
clarify the situation.  However, as argued earlier, if $R_b$ is the
only source of new physics, then the value of $\alpha_s(m_Z)$ deduced
from $\Gamma_{\rm had}$ will be lower than its SM-determined value, and
potentially in better agreement with the extrapolation from
lower energy data.  Furthermore,
$R_b$ is the most sensitive (among the partial $Z$-decay rates)
to new physics.  This is due, in part, to 
the large Higgs-top quark Yukawa coupling, which generates a
significant one-loop correction to $R_b$.   

Henceforth, I shall assume that $R_c$ is given by its Standard Model
prediction.  In the experimental determination of $R_b$, there
is some contamination of $c\bar c$ events in the $b\bar b$
sample that must be subtracted.  This subtraction depends on
the value of $R_c$ assumed.  Fixing $R_c$ to its Standard Model
value, a slightly smaller value of $R_b$ is found 
by the Electroweak working group compared to
the value quoted above:\cite{lepglobal}
\begin{eqnarray}
R_b=0.2205 \pm 0.0016,\qquad\qquad {\rm LEP/SLC~global~fit,}
\label{zbbnumber2}
\end{eqnarray}
roughly a three standard deviation discrepancy
from the Standard Model prediction.  

For completeness, I note here that
the $Z b \bar{b}$ vertex corrections can also affect the 
left-right $b \bar{b}$ asymmetry, 
${\cal{A}}_b \equiv (g_L^2 - g_R^2)/(g_L^2 +g_R^2)$,
where $g_L$ ($g_R$) are  the couplings of the left (right)
handed bottom quarks to the $Z$. The corrections to $R_b$ and 
${\cal{A}}_b$ can be parameterized as a function of the corrections
to the left-- and right--handed bottom quark vertices,\cite{abee}
\begin{equation}
\frac{\delta {\cal{A}}_b}{{\cal{A}}_b} = \frac{4 f_R f_L}
{f_L^4 - f_R^4} \left[f_R \delta g_L - f_L
\delta g_R \right],
\end{equation}
\begin{equation}
\frac{\delta R_b}{R_b} = \frac{2 \; (1-R_b)}{f_L^2 + f_R^2}
\left[f_R \delta g_R  + f_L \delta g_L \right],
\end{equation}
where $f_R=-\sin^2\theta_W/3$
and $f_L=1/2+f_R$ are the tree level
couplings of the right and left handed bottom quarks to
the $Z$. The dominant
top quark mass dependent one-loop $Zb\bar{b}$ vertex corrections
affect only the $Z$ coupling to the left--handed bottom quark,
$\delta g_L = -\alpha m_t/16 \pi \sin^2\theta_W$.
The large difference between the values of $f_L$ and $f_R$
implies that for $\delta g_R = 0$,
$\delta R_b/R_b \simeq 11.5 \; \delta {\cal{A}}_b/{\cal{A}}_b$.
Moreover, the current  determination of ${\cal{A}}_b$ at SLC 
is still subject to large experimental errors\cite{lepglobal}
\begin{eqnarray}
{\cal{A}}_b = \cases{0.841 \pm 0.053 &LEP/SLC global fit; \cr
0.935   ,&SM prediction.\cr}.
\label{Abnumber}
\end{eqnarray}
Therefore, ${\cal{A}}_b$ does not provide at present any significant
constraint on new physics beyond the Standard Model.

\section{The MSSM fit to precision electroweak data}

The Standard Model global fit to precision electroweak data of
Ref.~1 has a $\chi^2$ of 28 for 14 degrees of freedom,
which is not a very good fit to the data.  Of course, the goodness of
fit would improve significantly if the $R_c$ and/or $R_b$
measurements were not correct.  On the other hand, it is interesting to
examine whether any simple extension of the Standard Model can
dramatically alter the predicted values of $R_b$ without
seriously affecting the SM predictions for the other electroweak
observables.

In general, this is not an easy task.  For example, in some models 
that incorporate new physics beyond the Standard Model,
the effects of the new physics on precision electroweak
observables do not decouple in the limit where the scale of new physics 
becomes large compared to $m_Z$.  Such theories predict new
non-decoupling contributions to oblique radiative corrections (\ie,
corrections to gauge boson propagators), and
to vertex corrections such as the $Zb\bar b$ vector and axial vector
couplings.  Fits to the precision electroweak data which allow for
new physics contributions to the oblique corrections
find no evidence of any such effects.\cite{erler}  This imposes a strong
constraint on any model beyond the SM that attempts to improve the
goodness of the SM fit to the precision electroweak data.  Typically,
the existence of non-decoupling new physics worsens the global fit
(although, see Ref.~8 for an example where the global fit is improved).

The minimal supersymmetric extension of the Standard Model (MSSM) is
an example of a theory of decoupling new physics.  That is, if 
$M_{\rm SUSY}$ characterizes the scale of supersymmetric particle
masses, then the effects of virtual supersymmetric particle exchange
to $Z$ decay observables are suppressed by a factor of $m_Z^2/M_{\rm
SUSY}^2$.  If $M_{\rm SUSY}\gg m_Z$ (but we assume that $M_{\rm
SUSY}\lsim {\cal O}(1)$~TeV), then one remnant of the MSSM
exists below the scale $M_{\rm SUSY}$---a light CP-even Higgs boson
whose mass must be less than ${\cal O}(m_Z)$ [see Ref.~9
for an update on the light Higgs mass bound in the MSSM].   It follows
that if $M_{\rm SUSY}\gg m_Z$ (calculations\cite{decouple} 
show that it is sufficient to
have $M_{\rm SUSY}\gsim 200$~GeV), then the goodness of
the MSSM global fit to precision electroweak data is identical to that
of the SM global fit in the case of a light Higgs mass.  
  
If the MSSM global fit is to be better than the SM fit to precision
electroweak data, then the MSSM parameters must be such that not all
supersymmetric effects have decoupled.  In practice, this means that
some supersymmetric particle masses must be of ${\cal O}(m_Z)$ or
less.  This is good news for upcoming experimental searches at 
the LEP-2 and Tevatron colliders.  In particular, if the
discrepancies between precision electroweak observables and the SM
predictions are real and due to the effects of low-energy supersymmetry,
then some supersymmetric particles should be discovered during the
next few years.

\section{Low-energy supersymmetry and \protect{$\bold R_b$}}

Can parameters of the MSSM be chosen to improve the agreement between
theory and observation of $R_b$, while retaining the success of the SM
in describing the body of experimental electroweak data?\cite{BF}
[Since $R_c$
must be very close to the SM prediction in the MSSM, I shall take the
measured value of $R_b$ quoted in eq.~(\ref{zbbnumber2}).]  
During the past year, models of
low-energy supersymmetry have been examined in which $R_b$ is
slightly enhanced above the Standard Model 
prediction.\cite{rbmodels,topdec,wells,jellis}
In such models,
the global fit to the electroweak data is slightly improved.  Note that
in order to improve on the Standard Model fit, one must approximately
maintain the size of the Standard Model oblique corrections
while modifying the $Zb\bar b$ interaction.  In the MSSM, this
is possible if one takes large [small] values of the
mass parameters of the scalar super-partners of the
left [right] handed top quark, and small values of the Higgs superfield
mass parameter $\mu$.  Two distinct scenarios emerge depending on the
value of the parameter $\tan\beta$, the ratio of the two neutral Higgs
field vacuum expectation values.  For values of $\tan\beta\sim {\cal
O}(1)$, the dominant supersymmetric contribution to $R_b$ arises from
a one-loop triangle graph containing a light top-squark (dominantly
$\widetilde t_R$) and a light chargino (dominantly higgsino).
For values of $\tan\beta\sim m_t/m_b$, a new MSSM contribution
consisting of the triangle graph containing a light CP-odd Higgs
boson, $A^0$, plays a key role.  In the latter case, the enhanced
Higgs boson coupling to $b$-quarks when $\tan\beta\gg 1$ is the reason
for the enhanced value of $R_b$.

Although it was initially believed that $R_b$ could be as large as its
measured value [eq.~(\ref{zbbnumber2})] in the MSSM, recent
theoretical analyses suggest that this is unlikely.
In the MSSM, any physics leading to larger values of $R_b$
also contributes to non-standard top quark decays,
such as $t\to\widetilde t\widetilde\chi^0$ or
Based on the absence of
light charginos in the most recent LEP run at $\sqrt{s}=136$~GeV,
Ref.~15 quotes an absolute upper limit of $R_b<0.2174$
in the case of small $\tan\beta$.
In the large $\tan\beta$ regime, a rather light $A^0$ ($m_{A^0}\sim
40$~GeV) and a value of $\tan\beta\gsim 50$
is required to generate a large enough $R_b$.  However, in the MSSM, a
light $A^0$ implies a charged Higgs mass near its (approximate)
minimum value of $m_W$ (since $m_{H^\pm}^2\simeq m_W^2+ m_{A^0}^2$).
In this case, a recently derived 
$2\sigma$ upper bound,\cite{grossman} $\tan\beta<
41.6(m_{H^\pm}/m_W)$ is relevant.
Moreover, the low $m_{A^0}$, large $\tan\beta$ regime can be ruled
out due to the non-observation of $Z\to b\bar b A^0$ at LEP.  Ref.~14
concludes that a significantly enhanced $R_b$ in the large
$\tan\beta$ regime is ruled out.

I conclude this section with a brief description of a rather
unconventional low-energy supersymmetric model that does slightly
better in generating an enhanced value for $R_b$.  In the SM, $R_b$ is
suppressed relative to its tree-level prediction due to a negative
radiative corrections that grows quadratically with $m_t$.  Carena,
Wagner and I have constructed a four-generation low-energy
supersymmetric model in which $m_t\simeq m_W$.  In this model, the
effect of the top-quark radiative correction to $R_b$ is reduced.  We
find that $R_b\simeq 0.2184$, which is within one standard deviation
of the measured LEP value [eq.~(\ref{zbbnumber2})].  Moreover, with
this value of $R_b$, eq.~(\ref{newalpha}) implies $\alpha_s(m_Z) 
\simeq 0.112 \pm 0.005$, in good agreement with values of
$\alpha_s(m_Z)$ extrapolated from lower energy data.
Remarkably, such a four-generation model cannot
yet be excluded by present data.  In our model, $t\to \widetilde
t\widetilde\chi^0$ is the dominant decay, so that top quark decays contain few
hard leptons thereby eluding previous searches at hadron colliders.
The ``top-quark'' discovered at the Tevatron is the fourth generation
$t^\prime$ quark which decays dominantly into $bW^+$.  Finally, the
top quark mass deduced by the global fit of electroweak data can be
explained in our model as arising from the sum of oblique radiative
corrections generated by the third and fourth generation quarks and squarks.
However, such a model will be excluded if no light top squark 
is discovered in the 1996 LEP-2 run.  
Further details of this model can be found in Ref.~17.

\section{Conclusions}

If the anomalies in the $R_b$ and $R_c$ measurements persist, models
of low-energy supersymmetry will be hard-pressed to explain the
deviation from Standard Model expectations.  The discovery of new
physics beyond the Standard Model at LEP-2 and/or the Tevatron will be
essential for explaining the origin of the discrepancies.
On the other hand, if
the SM predictions for precision electroweak observables are
eventually confirmed, then 
new physics beyond the Standard Model must (almost certainly) be strongly
decoupled at energies of order $m_Z$.
The MSSM with heavy super-partner masses is
a model of this type; however, the
ultimate confirmation of such a picture will require the detection of
supersymmetric particles at future colliders such as the LHC.

\section*{Acknowledgments}

Much of the work reported here
is based on a collaboration with Marcela Carena and Carlos Wagner.
This work was supported in part by the U.S. Dept.~of Energy.
 

\clearpage
 
\end{document}